\documentclass[11pt,a4paper]{article}
\pdfoutput=1 
\usepackage{jheppub}
\usepackage{graphicx}
\usepackage{bm}
\usepackage{amsmath}
\usepackage{amsfonts}
\usepackage{braket}
\usepackage{amssymb}
\usepackage{epstopdf}
\usepackage{hyperref}
\usepackage{mathrsfs}
\usepackage{subfigure}
\usepackage{ulem}

\def\lsim{\mathrel{\rlap{\lower4pt\hbox{\hskip1pt$\sim$}}
    \raise1pt\hbox{$<$}}}    
\def\gsim{\mathrel{\rlap{\lower4pt\hbox{\hskip1pt$\sim$}}
    \raise1pt\hbox{$>$}}}                

\title{Baryogenesis, dark matter and inflation in the Next-to-Minimal Supersymmetric Standard Model}  
\author[a,b,c,d]{Csaba Bal\'azs,}
\author[e]{Anupam Mazumdar,}
\author[e]{Ernestas Pukartas}
\author[a,b,c]{and Graham White}

\affiliation[a]{School of Physics, Monash University, Victoria 3800, Australia}
\affiliation[b]{Monash Centre for Astrophysics, Monash University, Victoria 3800 Australia}
\affiliation[c]{ARC Centre of Excellence for Particle Physics, Monash University, Victoria 3800 Australia}
\affiliation[d]{Australian Collaboration for Accelerator Science, Monash University, Victoria 3800 Australia}
\affiliation[e]{Consortium for Fundamental Physics, Lancaster University, Lancaster, LA1 4YB, UK}

\emailAdd{csaba.balazs@monash.edu}
\emailAdd{a.mazumdar@lancaster.ac.uk}
\emailAdd{e.pukartas@lancaster.ac.uk}
\emailAdd{graham.white@monash.edu}

\abstract{Explaining baryon asymmetry, dark matter and inflation are important elements of a successful theory that extends beyond the Standard Model of particle physics. In this paper we explore these issues within Next-to-Minimal Supersymmetric Standard Model (NMSSM) by studying the conditions for a strongly first order electroweak phase transition, the abundance of the lightest supersymmetric particle (LSP), and inflation driven by a gauge invariant flat direction of MSSM - made up of right handed squarks. We present the regions of parameter space which can yield successful predictions for cosmic microwave background (CMB) radiation through inflation, the observed relic density for dark matter, and successful baryogenesis. Constrains by collider measurements (such as the recent Higgs mass bound), branching ratios of rare, flavour violating decays, and the invisible Z decay width are also imposed. We explore where dark matter interactions with xenon nuclei would fall within current bounds of XENON100 and 
the projected limits for the XENON1T and LUX experiments.} 
\begin{document}
\maketitle
\section{Introduction}
Beyond the Standard Model (BSM) particle physics should aspire to include the following three cosmological cornerstones. 

1. Inflation can explain the current temperature anisotropy in the cosmic microwave background (CMB) radiation measured by the 
Planck satellite~\cite{Ade:2013zuv}. Since inflation dilutes matter, the end of inflation must excite all the relevant Standard Model (SM) degrees of freedom without any excess of dark matter and dark radiation, along with the seed initial perturbations for the structure formation.
This can be achieved minimally by embedding inflation within a visible sector of BSM, such as the Minimal Supersymmetric Standard Model (MSSM) \cite{Allahverdi:2006iq, Allahverdi:2006cx,Allahverdi:2006we}. For a review on inflation, see~\cite{Mazumdar:2010sa}.

2. Dark matter is required to form the observed large scale structures of the Universe \cite{Peebles:1994xt}.  The lightest supersymmetric particle (LSP) is an ideal dark matter candidate since it is a weakly interacting massive particle (WIMP), and its interactions with the super-partners are sufficient to keep them in thermal bath until they decouple due to the expansion of the Universe and their abundance freezes out~\cite{Gondolo:1990dk}.  The thermal WIMP scenario is attractive due to its predictive power in estimating the abundance of the dark matter and potential link to the weak scale.  Within the MSSM the neutralino plays this role of dark matter ideally.  For reviews on dark matter, see~\cite{Jungman:1995df,Bertone:2004pz}.

3. Baryon asymmetry, at the observed level of one part in about $10^{10}$, is necessary to realize Big Bang nucleosynthesis ~\cite{Fields:2006ga}.  This asymmetry can be achieved by satisfying the three Sakharov conditions~\cite{Sakharov:1967dj}: baryon number ($B$) violation, charge ($C$) and/or charge-parity ($CP$) violation, and the out-of-equilibrium condition. These ingredients are present within the SM, but with a 126 GeV Higgs the model lacks sufficiently strong first order phase transition to keep the baryon asymmetry intact. For a review on baryon asymmetry, see~\cite{Rubakov:1996vz}.

The MSSM provides all three ingredients, however the electroweak baryogenesis in its context is becoming more and more constrained by the LHC data. 
In the simplest MSSM electroweak baryogenesis scenario (where the CP-violating phases catalyzing the asymmetry reside in the gaugino sector) it is getting hard, if not impossible, to generate enough asymmetry with a 126 GeV standard-like Higgs boson in a natural manner \cite{Carena:2012np}. 
So we turn our attention to the electroweak baryogenesis in the Next to Minimal Supersymmetric Standard Model (NMSSM).  (For a review on the NMSSM, see~\cite{Ellwanger:2009dp}.)  
The NMSSM has more flexibility as it introduces a new Standard Model gauge singlet which helps achieving a strongly first order electroweak phase transition, since the order parameter is now determined by the singlet sector and becomes essentially independent of the Standard Model-like Higgs mass. 

In this work, we present a highly efficient algorithm to delineate regions of the NMSSM with strongly first order phase transition.  After finding these regions we reject model points where the neutralino relic abundance exceeds the upper limit imposed by Planck by more than $3\sigma$. 
Some of the dark matter particles, in particular those with high singlino fraction, tend to escape the most stringent bounds given by the XENON100 \cite{Aprile:2012nq} and TEXONO~\cite{Li:2013fla} direct detection experiments, but they fall close to the regions where tantalizing positive signals from DAMA~\cite{Bernabei:2008yi}, CoGeNT~\cite{Aalseth:2010vx}, CRESST--II~\cite{Angloher:2011uu} and CDMS~\cite{Agnese:2013cvt} were announced.  The NMSSM singlet sector can also affect other properties of dark matter.  With a sizable singlino fraction the lightest neutralino can be very light even in the presence of heavy super-partners~\cite{Vasquez:2010ru,Boehm:2013qva}. 

The joint parameter space for satisfactory neutralino dark matter and first order electroweak phase transition has been studied in~\cite{Carena:2011jy}. 
Here we supersede that study by a new algorithm for finding regions where electroweak baryogenesis can be successful, by considering 
the parameter space for neutralino dark mater and successful inflation driven by the MSSM squarks, and updating all experimental constraints, especially the vital Higgs limits from the LHC~\cite{atlas-dec12,cms-dec12}.

In the next section we briefly summarize the relevant features of the NMSSM.  In section \ref{B-asymmetry} we examine the baryon washout condition and derive an algorithm to find regions of its parameter space satisfying a strongly first order electroweak phase transition. 
In section \ref{constraints app} we summarize the constraints applied to the theory, and in~\ref{scansss} we show model points jointly satisfying inflation, baryogenesis, and dark matter abundance. 
We discuss implications on inflation in \ref{inflationnn} and finally conclude. 


\section{The Next-to-Minimal Supersymmetric Standard Model}\label{NMSSM-intro}

In this work we consider the $Z_3$ conserving, scale invariant, version of the the Next-to-Minimal Supersymmetric Standard Model \cite{Fayet:1974pd, Nilles:1982dy} which is defined by the superpotential (see, for instance, Ref. \cite{Ellwanger:2009dp})
\begin{equation} 
 W = W^{\rm MSSM}|_{\mu=0} + \lambda \widehat{S} \widehat{\bf H}_u \widehat{\bf H}_d + \frac{1}{3} \kappa \hat{S}^3 . 
\end{equation}
Above $W^{\rm MSSM}|_{\mu=0}$ is the MSSM superpotential without the $\mu$ term (as defined in Ref. \cite{Balazs:2009su}) and $\widehat{\bf H}_{u,d}$ ($\widehat{S}$) are SU(2) doublet (singlet) Higgs superfields: $\widehat{\bf H}_u = (\widehat{H}^+_1,\widehat{H}^0_1)^T$, $\widehat{\bf H}_d = (\widehat{H}^0_2, \widehat{H}^+_2)^T$.  The superscript in $\widehat{H}_i^{\pm,0}$ denotes the electric charge of the component.
The corresponding soft supersymmetry breaking scalar potential
\begin{eqnarray}
 V_{\rm soft} &=& V_{\rm soft}^{\rm MSSM}|_{B = 0} + V_{\rm soft}^{\rm NMSSM}
\end{eqnarray}
contains the MSSM soft terms with $B$ set to zero \cite{Balazs:2009su} and
\begin{eqnarray}
 V_{\rm soft}^{\rm NMSSM} = m_s^2 |S|^2 - \lambda A_\lambda S {\bf H}_u {\bf H}_d + \frac{1}{3} \kappa A_\kappa S^3 + h.c.
\end{eqnarray}
Here ${\bf H}_{u,d}$ and $S$ denote the scalar components of the neutral Higgs superfields.  During electroweak symmetry breaking the neutral components of these will acquire a non-zero vacuum expectation value.
The MSSM terms above are \cite{Balazs:2009su}
\begin{eqnarray} 
 V_{\rm soft}^{\rm MSSM}|_{B=0} &=& V_{\rm soft H}^{\rm MSSM} + V_{\rm scalar} + V_{\rm gaugino} + V_{\rm tri} .
\end{eqnarray}

It is important for electroweak baryogenesis that the Higgs potential
\begin{eqnarray} 
 V = V_D + V_{\rm soft H}^{\rm MSSM} + V_{\rm soft}^{\rm NMSSM} + V_{\rm H}^{\rm NMSSM} ,
\end{eqnarray}
receives contributions both from the MSSM
\begin{eqnarray} 
 V_D = \frac{g_1^2+g_2^2}{8}({\bf H}_u^2-{\bf H}_d^2)^2 ,
\end{eqnarray}
\begin{eqnarray} 
 V_{\rm soft H}^{\rm MSSM} = m_{H_u}^2 |{\bf H}_u|^2 + m_{H_d}^2 |{\bf H}_d|^2 ,
\end{eqnarray}
and the NMSSM
\begin{eqnarray} 
 V_{\rm H}^{\rm NMSSM} &=& \lambda^2 |S|^2 (|{\bf H}_u|^2+|{\bf H}_d|^2) + \lambda^2 |{\bf H}_u {\bf H}_d|^2 + \kappa^2 |S|^4 \nonumber \\
                   &&+~ \kappa \lambda S^2 {\bf H}_u^*{\bf H}_d^* + h.c.
\end{eqnarray}
While $\lambda$ is a free parameter, perturbativity up to the Grand Unification Theory scale restricts it below 0.7.  In this work we respect this limit and do not take $\lambda$ higher.
The cubic singlet coupling $\kappa$ breaks the global U(1) Peccei-Quinn symmetry \cite{Jeong:2011jk}, and when $\kappa$ vanishes this symmetry is restored.  For simplicity, in this work we consider the limit where $\kappa$ is small, that is the Peccei-Quinn limit.  This limit is also motivated by the desire to obtain a light dark matter candidate, as indicated in the next paragraph.

In this work we the lightest neutralino is the dark matter candidate.  Neutralinos in the NMSSM are admixtures of the fermionic components of five superfields: the U(1)$_Y$ gauge boson ${\widehat B}$, the neutral component of the SU(2)$_L$ gauge boson ${\widehat W_3}$, the neutral components of each Higgs doublet ${\widehat H}_u$ and ${\widehat H}_d$, and the singlet ${\widehat S}$.  The mass of the lightest neutralino originates from soft supersymmetry breaking
\begin{eqnarray} 
 V_{\rm gaugino} &=& \frac{1}{2} (M_1 \bar{\tilde{B}} \tilde{B} + M_2 \bar{\tilde{W}}_i \tilde{W}_i + M_3 \bar{\tilde{G}}_a \tilde{G}_a) ,
\end{eqnarray}
where the fields above are the fermionic components of the vector superfields.  The admixture of the lightest neutralino is controlled by the neutralino mass matrix which, in the $(-i \tilde{B}, -i \tilde{W}_3, \tilde{H}_d, \tilde{H}_u, \tilde{S})$ basis, is given by the symmetric matrix \cite{Ellwanger:2009dp}
\begin{eqnarray} 
 \mathcal {M}_{\widetilde{\chi}} = 
 \left( \begin{array}{ccccc}
  M_1 &   0 & -\frac{g_1 \langle H_d \rangle}{\sqrt{2}} &~~\frac{g_1 \langle H_u \rangle}{\sqrt{2}} &                            0\\
      & M_2 &~~\frac{g_2 \langle H_d \rangle}{\sqrt{2}} & -\frac{g_2 \langle H_u \rangle}{\sqrt{2}} &                            0\\
      &     &                                         0 &                -\lambda \langle S \rangle & -\lambda \langle H_u \rangle\\
      &     &                                           &                                         0 & -\lambda \langle H_d \rangle\\
      &     &                                           &                                           &  ~\kappa \langle S \rangle
 \end{array} \right) .
\end{eqnarray}
Here $g_{1,2}$ are the U(1)$_Y$ and SU(2)$_L$ gauge couplings, and $\langle X \rangle$ denote vacuum expectation values.  To obtain the mass eigenstates, we have to diagonalize the neutralino mass matrix. 
This can be done with the help of a unitary matrix $N_{ij}$ whose entries provide the mixing amongst gauginos, higgsinos and the singlino.  The lightest neutralino, for example, is given by:
\begin{equation}
\label{mixinggg}
\widetilde{\chi}_1^0=N_{11}\widetilde{B}+N_{12}\widetilde{W}^0+N_{13}\widetilde{H}_d^0+N_{14}\widetilde{H}_u^0+N_{15}\widetilde{S}, 
\end{equation}
where $\lvert N_{11}\rvert^2$ gives the bino, $\lvert N_{12}\rvert^2$ the wino, $\lvert N_{13}\rvert^2+\lvert N_{14}\rvert^2$ the higgsino and $\lvert N_{15}\rvert^2$ the singlino fraction.  When, for example, $\kappa \langle S \rangle$ is the smallest entry of the mass matrix, the lightest neutralino tends to be singlino dominated.  Alternatively, a small $M_1$ entry can render the lightest neutralino to acquire mostly bino admixture. 

\section{Baryon asymmetry}
\label{B-asymmetry}

As mentioned above, three conditions have to be met to generate baryon asymmetry: $B$ violation, $C$ and/or $CP$ violation, and departure from thermal equilibrium.  Remarkably, these conditions can be met in the Standard Model of particle physics.  The difference of the baryon and lepton numbers, $B-L$, is an exactly conserved quantity in the SM (and within NMSSM).  While at low temperatures $B$ and $L$ are individually conserved with a good approximation, at very high temperatures baryon number violation in unsuppressed through sphaleron processes \cite{Rubakov:1996vz, earlyUniverseanomalies}.  Unfortunately, there is not enough $CP$ violation in the standard CKM matrix to generate the observed baryon asymmetry.  In the NMSSM, sufficient amount of $CP$ violation can be added to the gaugino and singlino sectors.  Provided that the third Sakharov condition is met, it is possible to generate the observed asymmetry in the NMSSM.  In this paper we focus on the latter: Where in the NMSSM parameter space a strong first order electroweak phase transition be achieved?

At very high temperatures electroweak symmetry is restored \cite{hightemphiggs}. As the Universe cools, the electroweak symmetry is broken during the phase transition. If the phase transition is first order, the  electroweak symmetry is broken through tunneling processes.  This occurs when the effective potential has degenerate minima $\{0, \varphi _c\}$, where $\varphi$ denotes the Higgs vacuum expectation value (VEV). In such a case bubbles of broken phase grow in the otherwise symmetric vacuum, until the phase transition is complete and symmetry is broken everywhere in the Universe. Within the symmetric phase sphaleron processes , and therefore baryon violating processes are unsuppressed whereas they are exponentially dampened within the broken phase. The  $C$ and $CP$ violating processes near the bubble walls can create a large baryon asymmetry. 
If the phase transition is strongly first order the baryon asymmetry will be preserved \cite{Rubakov:1996vz}. 
To this end we require \cite{baryonwashout1,baryonwashout2,baryonwashout3,
baryonwashout4}
\begin{equation} 
 \frac{T _c}{\varphi_c} \equiv \gamma \lesssim 1 ,
\end{equation}
where $T_c$ is the temperature at which the effective potential obtains degenerate minima. This is known as the baryon washout condition.  The Standard Model cannot satisfy the baryon washout condition for a Higgs like particle of mass about 125 GeV \cite{BAinSM}. This experimental constraint also all but rules out electroweak baryogenesis in the MSSM \cite{csaba}. Recent work however has suggested that the NMSSM is compatible with a strongly first order phase transition and the observed Higgs mass \cite{wagner}. As noted before the baryon washout condition is not the only hurdle that prevents the Standard Model from being consistent with electroweak baryogenesis, there is also the issue of insufficient $CP$ violation. Here we focus on the baryon washout condition and postpone the detailed investigation of the relevant $CP$ violating phases in the NMSSM to a later work.

\subsection{The scalar potential at high temperature}
\label{athightemp}
Previous analysis of the electroweak phase transition within the NMSSM near the Peccei-Quinn limit found that the parameter space that satisfies the baryon washout condition is heavily constrained \cite{wagner}. Seeing as we also wish to satisfy other, in some cases rather strict, experimental and cosmological constraints, we are motivated to find a numerically efficient way of finding regions of parameter space that allow for a strongly first order phase transition. In a prior analysis, Wagner $et \ al$. considered a toy model which included the tree level effective potential of the NMSSM at the Peccei-Quinn limit with the largest temperature corrections. Our strategy will be to derive a semi-analytic solution to the toy model from  Ref.~\cite{wagner} and consider higher order temperature corrections, loop corrections and small deviations from the so called Peccei-Quinn limit as perturbations to the toy model solution. If we only scan regions of parameter space where our approximations hold, it 
is numerically efficient to produce a very 
large volume of points that easily satisfy all of our cosmological and experimental constraints.

The one loop temperature corrections to the effective potential are:
\begin{equation} V(T,m , \pm 1) = g_{\pm}  \frac{T^4}{2 \pi ^2} \int _0 ^\infty dx  x^2\log \left( 1 \pm e^{-\sqrt{x^2 + \frac{m(\phi)^2 }{T^2}}} \right) \,,
\end{equation}
where $g_\pm $ is the number of fermionic or bosonic degrees of freedom respectively. Similarly the argument is $+$ for fermions and $-$ for bosons. The high temperature expansion is up to an overall temperature dependent constant \cite{jackiw}
\begin{eqnarray} V(T,m,+1) &\sim& \frac{g_+m^2(\phi )T^2}{24} -\frac{g_+[m(\phi)^2]^{3/2}(\phi )T}{12 \pi}-\frac{g_+m(\phi) ^4 }{64\pi^2} \log \left( \frac{m(\phi)^2}{a_bT^2} \right) \nonumber \\ & \equiv & \frac{g_+m^2(\phi )T^2}{24} -\frac{g_+[m(\phi)^2]^{3/2}(\phi )T}{12 \pi} + \Delta V_T \,,\nonumber \\  \end{eqnarray}
for bosons, and
\begin{eqnarray} V(T,m,- 1) &\sim & \frac{g_-m^2(\phi )T^2}{48} +\frac{g_-m(\phi) ^4 }{64\pi^2} \log \left( \frac{m(\phi)^2}{a_fT^2} \right) \nonumber \\ & \equiv & \frac{g_-m^2(\phi )T^2}{48} + \Delta V_T\,, \end{eqnarray}
for fermions. Here $a_b=(4 \pi e^{-\gamma _E})^2$ and $a_f=( \pi e^{-\gamma _E})^2$. We have also identified the $\log$ term as $\Delta V_T$ to highlight that these terms will be treated as a perturbation. To keep notation compact, all small temperature corrections that are not included in our toy model (i.e. the ones that we treat as a perturbation) and their sum we denote as $\Delta V _T$. It should be noted that the high temperature approximation for the one loop temperature corrections cannot be assumed to be valid. Indeed it is only valid when $m(\phi )/ T \lesssim 2.2$ and $1.9$ for bosons and fermions respectively. Our numerical scans stay away from this limit, we generally have $m(\phi )/ T \lesssim 1.5$. 
The temperature dependent effective potential is a function of the Higgs field and the singlet field which we denote $\varphi _S$. (We also use the short hand that $\varphi \equiv \sqrt{ H_u^2 + H_d^2}$.) Within one loop accuracy under the high temperature expansion, it is given by
\begin{equation} V(\varphi , \varphi _s, T)  = V^{\rm T} + \Delta V_S + \Delta V _{\rm loop} +\Delta V_T \  . \end{equation}
Here we have defined $\Delta V_{\rm loop}$ as the loop corrections, $\Delta V_S$ as the terms that violate the 
Peccei-Quinn limit (which is approximately $ \kappa A_\kappa \varphi _s^3/3$).  The term $V^{\rm T}$ is the toy model effective potential from Ref.~\cite{Carena:2011jy}, which is given by
\begin{eqnarray} V^{\rm T} = M^2\varphi ^2 +cT^2\varphi ^2 - ET\varphi ^3 +m_s^2\varphi _s^2 + \lambda ^2 \varphi _s^2 \varphi ^2 -2 \tilde{a} \varphi ^2 \varphi _s+\frac{\tilde{\lambda}}{2} \varphi ^4 \ , \end{eqnarray}  
where
\begin{eqnarray}   M^2&=& m_{H_u}^2 \cos ^2 \beta + m _ {H_d}^2 \sin ^2 \beta , \nonumber \\ 
\tilde{a} &=& \lambda A_\lambda \sin \beta \cos \beta , \nonumber \\  
\frac{\tilde{\lambda}}{2} &=&\frac{g_1^2+g_2^2}{8}\left( \cos ^2 \beta - \sin ^2 \beta \right)^2 +\lambda ^2 \sin ^2 \beta \cos ^2\beta  +\frac{\delta \tilde{\lambda}}{2} . \nonumber \\  \end{eqnarray}
In the last equation the parameter $\delta \tilde{\lambda}$ acquires large loop corrections from the stop mass. 
Note that we have not included any temperature corrections to the bare mass of the singlet. Therefore our scheme is valid in the region where 
\begin{equation} \frac{\lambda \varphi _s }{\gamma \varphi _c} \gtrsim 1.9 \ . \end{equation}
 In the region where temperature corrections to the bare singlet mass become important it is possible for a first order phase transition to occur when $m_s^2<0$. These points however are relatively rare and we lose little in ignoring them.

We recall that at the critical temperature the effective potential obtains degenerate minima with one minima at $\varphi = 0$. It is easy to see that $V(0,0,T)=V^{\rm T}(0,0,T)=0$. Let the critical temperature and the non zero VEV at this temperature for our toy model be denoted by $T_c$ and $\varphi _c$, respectively.  It is also useful to define $\tilde{\gamma} = T/\varphi$ for $T\neq T_c$. Note that $V$ can be written as a function of $\varphi$,  $\varphi _S$ and $\tilde{\gamma}$. We will denote the fields, $\varphi _x$, away from the respective minima as $\tilde{\varphi }_x$. Let us assume that $V$ is continuous in its three arguments near the critical temperature. It is then apparent that $V(\varphi  _c + \delta \varphi _c, \tilde{\varphi _s} + \delta \varphi _s, \gamma + \delta \gamma ) = 0$, where the non trivial VEV of the full temperature dependent potential at the critical temperature is a small perturbation to the tree level critical VEV, $\varphi _c + \delta \varphi _c$. Similarly, 
the singlet VEV at the critical and the inverse order parameter both obtain small corrections, $\delta \varphi _s, \delta \gamma$, respectively. From the small change formula in three variables, we can write:
\begin{eqnarray} &&V(\varphi _c + \delta \varphi _c, \tilde{\varphi _s} + \delta \tilde{\varphi _s}, \gamma + \delta \gamma ) \nonumber \\ &\approx& V^{\rm T}(\varphi _c , \tilde{\varphi }_s, \gamma )  + \left. \left(\Delta V _T + \Delta V_{\rm loop} + \Delta V_S \right) \right|_{\varphi _c, \tilde{\varphi _s}, \gamma} \nonumber \\ &&+ \left. \frac{\partial V ^{\rm T}}{\partial \tilde{\varphi} } \right|_{\varphi _c, \varphi _s, \gamma}\delta \varphi _c +\left. \frac{\partial V ^{\rm T}}{\partial \tilde{\varphi} _s } \right|_{\varphi _c, \varphi _s, \gamma} \delta \varphi _s +\left. \frac{\partial V ^{\rm T}}{\partial \tilde{\gamma} } \right|_{\varphi _c, \varphi _s, \gamma} \delta \gamma \ . \nonumber \\ \end{eqnarray}
The first term on the right hand side of the above equation is identical to zero for the reasons discussed above. Furthermore, the derivative of our toy model effective Lagrangian with respect to either $\varphi $ or $\varphi _c$ is also zero by definition when the derivative is evaluated at its minimum. Setting the left hand side of the above equation to zero and defining $\Delta V \equiv \Delta _T + \Delta _S + \Delta _{\rm loop}$, we can then write:
\begin{equation} \delta \gamma = -\Delta V  \left. \left/ \frac{\partial V^T}{\partial \tilde{\gamma}} \right. \right|_{\varphi _c, \varphi _s , \gamma} \ . \end{equation}
Noting that $\partial \varphi / \partial \tilde{\gamma } =- \varphi/\tilde{\gamma}$, we can write
\begin{eqnarray}  
\frac{\partial V}{\partial \tilde{\gamma}} &=& \frac{2 G \varphi ^2}{\tilde{\gamma}} - 2 c \tilde{\gamma } \varphi ^4 -\frac{2 \tilde{\lambda} \varphi ^4}{\tilde{\gamma}} + 3E \varphi ^4 + \frac{\tilde{a}^2 \varphi ^4 (4m_s^2 + 2 \lambda ^2 \varphi ^2)}{\tilde{\gamma}(m_s^2+\lambda ^2 \varphi ^2)^2} .\end{eqnarray}
Finally we solve our toy model.  We begin this calculation by insisting that the zero temperature VEV is $v = 174$ GeV. This gives us the relation:
\begin{equation} -M^2 = v^2\left(\tilde{\lambda} - \frac{\tilde{a}^2(2m_s^2+\lambda^2v^2)}{(m_s^2+\lambda^2v^2)^2}\right) \equiv G \ . \end{equation}
Using the condition of degenerate minima occurring at a critical temperature, it is easy to derive the following equation
\begin{eqnarray}0 =-\frac{\tilde{\lambda}}{2}+\gamma E-c\gamma^2+ \frac{\sqrt{\tilde{a}^2(\tilde{\lambda}-\gamma E)}}{ \sqrt{2} m_s} + \frac{\lambda^2 G}{-m_s^2 + \sqrt{\frac{2 \tilde{a}^2 m_s^2}{\tilde{\lambda}-\gamma E}}}   \equiv  F(\gamma) \ . 
\label{F-gamma-0}
\end{eqnarray} 
The details of this calculation are given in the appendix. Note that, apart from $\gamma$, this equation is a function of only four parameters: $\{ m_s, \lambda, A_\lambda, \tan \beta\}$. We therefore calculate $\delta \gamma $ for values of  $\{ m_s, \lambda, A_\lambda, \tan \beta\}$ such that $F(\gamma + \delta \gamma)$ is significantly smaller than any of its five components. We ensure that all components of $\Delta V $ are small compared to the derivative of $V^{\rm T}$ with respect to gamma evaluated at the vev at the critical temperature. Finally we insist that $\delta \gamma \lesssim 0.4$. The baryon washout condition is satisfied for $\gamma + \delta \gamma \lesssim 1$.

\section{Constraints applied}
\label{constraints app}
To find regions consistent with experiment we used \texttt{micrOMEGAs2.4}~\cite{Belanger:2010gh} coupled to NMSSMTools~\cite{Ellwanger:2005dv} to calculate observables and performed a scan over NMSSM parameter ranges shown in Table~\ref{tab:summary}.  Most importantly, we require the dark matter relic abundance to be consistent with Planck along with the constraint of strongly first order electroweak phase transition of Eq.~(\ref{F-gamma-0}).  Additionally, we impose current limits from various experiments, as we enumerate them below.

\begin{itemize}
 \item {{\it Relic abundance}: We require that model points satisfy an upper limit on dark matter relic abundance observed by the Planck satellite, i.e. $\Omega_{{\widetilde{\chi}}_1^0} h^2<0.128$ \cite{Ade:2013zuv}.  We find that in large part of the parameter space the  
 lightest neutralino is not enough to account for the total Planck measured value, and multi-component dark matter should be considered.
The points which pass the Planck constraint within $3\sigma$ CL i.e. $0.1118<\Omega_{{\widetilde{\chi}}_1^0} h^2<0.128$ \cite{Ade:2013zuv} are highlighted.}

 \item {{\it Higgs mass}: We impose the  LHC bound on the Higgs boson mass by taking the  
 combined theoretical and experimental uncertainties within the following range, i.e. $121.5<m_h<129.5$ GeV.}

 \item {{\it Direct dark matter detection}: We illustrate the bounds on a neutralino--nucleon interaction cross--section as measured by the XENON100 experiment~\cite{Aprile:2012nq} and CRESST-II~\cite{Angloher:2011uu}. We also consider the projected bounds of XENON1T~\cite{Aprile:2012zx} and LUX~\cite{Akerib:2012ys}. These bounds are derived using standard assumptions, i.e. dark matter density in the Galactic halo 
 $\rho_{\widetilde{\chi}_1^0}=0.3$ GeV/cm$^3$, the circular velocity $v=220$ km/s and the Galactic escape velocity $v_{esc}=544$ km/s ~\cite{Smith:2006ym}.}

 \item{{\it Flavour physics}:  We enforce limits on the branching ratios of flavour violating decays, $\mathcal{B}(B_s \rightarrow \mu^+\mu^-)= \left(3.2^{+1.5}_{-1.2}\right)\times 10^{-9}$ \cite{lhcb} and $\mathcal{B}(b \rightarrow s\gamma)=(3.55\pm 0.26)\times10^{-4}$ \cite{Asner:2010qj}.  Baryogenesis does not conflict with these constraints, since these processes are enhanced for large values of $\tan{\beta}$ while electroweak baryogenesis in the NMSSM has a preference for moderate $\tan \beta$ values \cite{Carena:2011jy}.}

 \item {{\it Muon anomalous magnetic moment}:  We require the supersymmetric contribution to $g_\mu -2$ to be in the range: $-2.4\times10^{-9}<\delta a_{\mu}^{SUSY}<4.5\times10^{-9}$ \cite{Arbey:2011un}. }

 \item {{\it Chargino mass}: We take the lower LEP bound on a mass of chargino to be $m_{\widetilde{\chi}^+_1}>103.5$ GeV \cite{chargino}.\footnote{We used a general limit on the chargino mass, however for possible caveats, see Ref. \cite{Abdallah:2003xe}.}
 A null result in LEP searches on a process $e^+e^-\rightarrow\widetilde{\chi}_1\widetilde{\chi}_j$ 
 with $j>1$, sets an upper bound on the neutralino production cross--section $\sigma(e^+e^-\rightarrow\widetilde{\chi}_1\widetilde{\chi}_j)\lesssim10^{-2} \text{pb}$ \cite{Ellwanger:2009dp}, which can be translated into $(m_{\widetilde{\chi}^0_1}+m_{\widetilde{\chi}^0_2})>209$ GeV~\cite{Menon:2004wv}. This limits the mass of the lightest neutralino from below.}

 \item{{\it Invisible $Z$ boson decay width}:  In the light neutralino regions where $m_{\widetilde{\chi}_1^0}<M_Z/2$, one has to take into account of the invisible decay width of Z boson into neutralinos.  Analytical expression for this process is given by \cite{Ellwanger:2009dp}:
 \begin{equation}
  \label{invisible}
  \Gamma_{Z\rightarrow\widetilde{\chi}_1^0\widetilde{\chi}_1^0}=\frac{M_Z^3G_F}{12\sqrt{2}\pi}(N_{13}^2+N_{14}^2)^2\bigg(1-\frac{4m_{\widetilde{\chi}^0_1}^2}{M_Z^2}\bigg)^{3/2}\,,
\end{equation}
where $N_{13}^2$ and $N_{14}^2$ are Higgsino fractions coming from Eq.~(\ref{mixinggg}). This relationship is derived assuming 
three massless neutrinos. In order to satisfy this constraint the lightest neutralino must mostly be either a bino, a wino, or a singlino with minimal or no admixture from higgsinos. }

\end{itemize}

Our constrains are summarized in Table.~\ref{tablecons}.


\begin{table}[ht!]
\begin{center}
\begin{tabular}{|c|c|c|}
\hline 
  Quantity & Value & Source\\ 
\hline \hline 
 $\Omega_{{\widetilde{\chi}}_1^0} h^2$ & $<0.128$ & \cite{Ade:2013zuv} \\ \hline
       $m_h$  & $125.5\pm 4$ GeV & \cite{atlas-dec12, cms-dec12} \\ \hline
       $(m_{\widetilde{\chi}^0_1}+m_{\widetilde{\chi}^0_2})$ & $>209$ GeV    & \cite{Ellwanger:2009dp} \cite{Menon:2004wv}  \\ \hline
       $\mathcal{B}(B_s \rightarrow \mu^+\mu^-)$ & $ \left(3.2^{+1.5}_{-1.2}\right)\times 10^{-9}$ & \cite{lhcb}\\ \hline
       $\mathcal{B}(b \rightarrow s\gamma)$ &  $(3.55\pm 0.26)\times10^{-4}$ & 
\cite{Asner:2010qj} \\ \hline
       $\delta a_\mu$ & $ (-2.4:4.5) \times10^{-9}$ & \cite{Arbey:2011un} \\ \hline
       $\Gamma_{Z \rightarrow \widetilde{\chi}^0_1\widetilde{\chi}^0_1}$ & $ <3 \text{MeV}$ & \cite{lepinvisible} \\ \hline
       $m_{\widetilde{\chi}^+_1}$ & $>103.5$ GeV   & 
\cite{chargino} \\ \hline
\end{tabular}
\caption{List of the experimental constraints which we imposed in our NMSSM scan.}
\label{tablecons}
\end{center}
\end{table}

\begin{table}[ht!]
\begin{center}
\begin{tabular*}{0.519\textwidth}{|c|c|}
\hline 
  Parameter& Range \\ 
\hline 
\hline 
       $\lambda$  & $[0:0.7]$ \\ \hline
       $A_\lambda$ & [0:10000] GeV\\ \hline
       $\tan\beta$ & [1:65] \\ \hline
       $\mu$ & $[0:A_\lambda\cos\beta\sin\beta]$ GeV\\ \hline
       $A_\kappa\cdot \kappa$ & $[0:0.01]$ GeV\\ \hline
       $M_1$ & [10:3000] GeV\\ \hline
       $M_2$ & [10:4000] GeV\\ \hline
       $M_3$ & [800:6000] GeV\\ \hline
       $m_{\widetilde{e}_L}=m_{\widetilde{\mu}_L}=m_{\widetilde{\tau}_L}$ & 6000 GeV\\ \hline
       $m_{\widetilde{e}_R}=m_{\widetilde{\mu}_R}=m_{\widetilde{\tau}_R}$ & 6000 GeV\\ \hline
       $m_{\widetilde{Q}_{1L}}=m_{\widetilde{Q}_{2L}}$ & 6000 GeV\\ \hline
       $m_{\widetilde{Q}_{3L}}$ & 3300 GeV\\ \hline
       $m_{\widetilde{u}_R}=m_{\widetilde{c}_R}$ & 6000 GeV\\ \hline
       $m_{\widetilde{t}_R}$ & 4000 GeV\\ \hline	
       $m_{\widetilde{d}_R}=m_{\widetilde{s}_R}=m_{\widetilde{b}_R}$ & 6000 GeV\\ \hline
       $A_t$  & -5000 GeV\\ \hline
       $A_\tau$  & -2500 GeV\\ \hline
       $A_b$  & -2500 GeV\\ \hline
\end{tabular*}
\caption{Scan ranges and fixed values of the NMSSM parameters.  The upper bound on $\mu$ comes from our requirement for $m_s^2$ to be positive.  The first four parameters are constrained by the strongly first order phase transition.  We vary the gaugino masses in order to explore the dark matter phenomenology and inflation, which we shall discuss below.}
\label{tab:summary} 
\end{center}
\end{table}

\subsection{Scanning the parameter space for dark matter and first order phase transition} 
\label{scansss}
Table.~\ref{tab:summary} shows the parameter ranges of our scan.  
We fixed the first and second generation sfermionic masses to high values in order to avoid large potential suspersymmetric contributions to electron and nuclear electric dipole moments~\cite{EDMs}.  
The masses of left and right handed stops are adjusted to yield the measured value of the Higgs boson mass.  We varied the electroweak gaugino masses in a wide range to explore dark matter phenomenology. 
Varying the gluino mass is important to determine the running of the  $\widetilde u \widetilde d \widetilde d$ inflaton.  
The selected ranges also guarantee that our spectrum does not conflict with the LEP~\cite{neutralino2}, ATLAS~\cite{atlas-susy} and CMS~\cite{cms-susy} bounds on squarks and sleptons. 


\begin{figure}[t]
\centering
\includegraphics[width=0.495\linewidth]{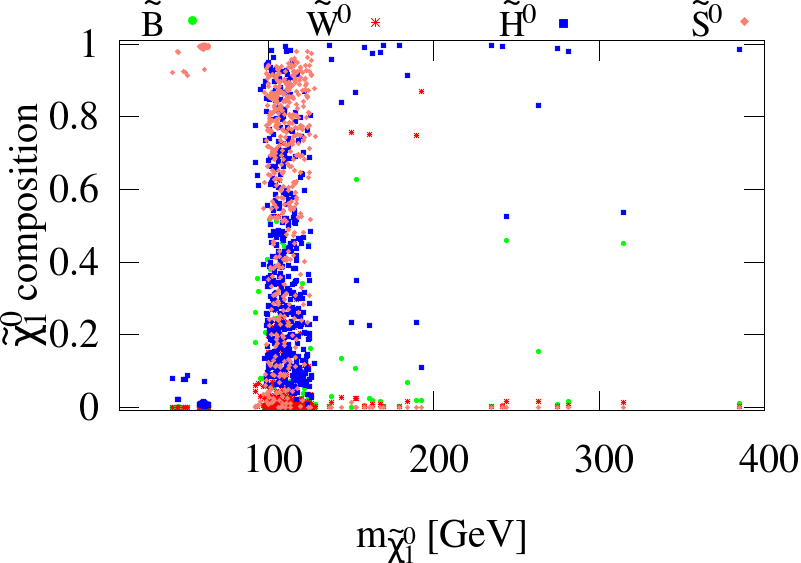}
\includegraphics[width=0.495\linewidth]{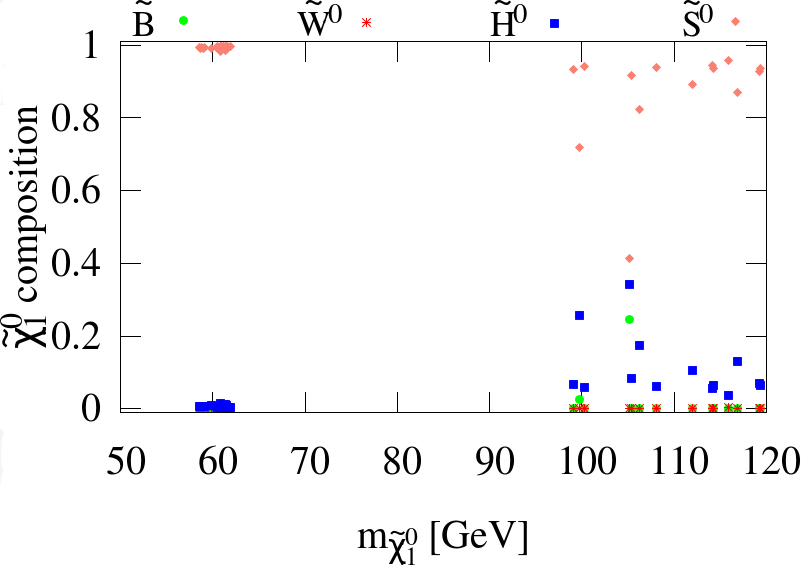}
\caption{Bino (green dots), wino (red stars), higgsino (blue squares), and singlino (pink diamonds) components of the lightest neutralino for scanned model points.  All points shown satisfy the condition of the strongly first order electroweak phase transition and pass all constraints listed in section \ref{constraints app}.  On the right hand panel we only show the points that fall within $3\sigma$ of the Planck central value for the relic abundance of dark matter: $0.1118<\Omega_{{\widetilde{\chi}}_1^0} h^2<0.128$ \cite{Ade:2013zuv}.}
\label{composition}
\end{figure}


Fig.~\ref{composition} shows the bino (green dots), wino (red stars), higgsino (blue squares), and singlino (pink diamonds) fractions of the lightest neutralino.  
We restrict the relic density of the lightest neutralino below $\Omega_{CDM} h^2<0.128$ which is the upper value on the dark matter abundance set by Planck at $3\sigma$ confidence level.
As mentioned above, neutralinos with masses $m_{\widetilde{\chi}_1^0}<m_Z/2$ have to have small higgsino fraction due to strict limits on the invisible Z boson decay (from Eq.~(\ref{invisible})) and the mass of the lightest chargino.  Also, light dark matter regions are very fine tuned and require separate detailed analysis and more sophisticated scanning techniques, see for instance, Ref.~\cite{Boehm:2013qva}. 
  
The heavier neutralino can have a larger higgsino fraction. 
This is also connected to the fact that the positivity of  $m_s^2$ sets an upper bound on the parameter $\mu<A_\lambda\cos{\beta}\sin{\beta}$, which allows larger higgsino fraction in the lightest neutralino. 
As we shall argue in the next paragraph this opens up more annihilation channels to satisfy the relic density constraint. 
 

\begin{figure}[t]
\centering
\includegraphics[width=0.75\linewidth]{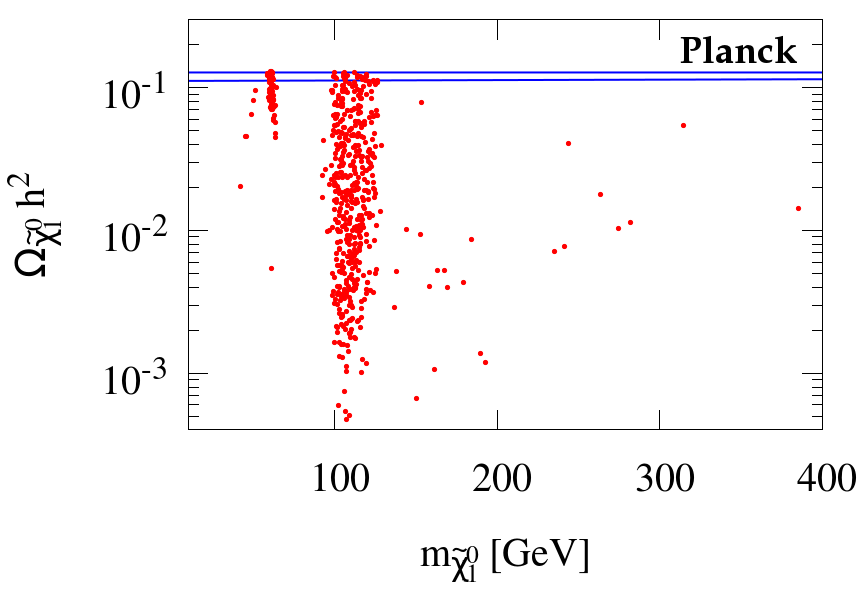}
\caption{Relic abundance versus the mass of the dark matter particle for our scan.  
The blue lines indicate the bounds on dark matter density implied by the Planck satellite: $0.1118<\Omega_{{\widetilde{\chi}}_1^0} h^2<0.128$.
All the points satisfy the condition for a first order electroweak phase transition, as indicated by Eq.~(\ref{F-gamma-0}).  
They also pass all the constraints listed in section \ref{constraints app}.}
\label{omega}
\end{figure}


In Fig.~\ref{omega}, we show the dark matter relic density dependence on $m_{\widetilde{\chi}_1^0}$.  
We only show points for which the dark matter relic density to falls below the upper limit from Planck, that is to satisfy $\Omega_{\widetilde{\chi}_1^0} h^2<0.128$ at $3\sigma$ confidence level \cite{Ade:2013zuv}. 
The peculiar clustering of the points in this plot is understood as follows.
The first, smaller group around $m_{\widetilde{\chi}_1^0}\approx 63$ GeV is due to neutralino annihilation through the 126 GeV Higgs.
This resonant annihilation depletes the neutralino abundance making it possible to satisfy the Planck bound.
The second, larger group of points originate from the lightest neutralinos with an enhanced higgsino  component coupling to $Z$ boson.
As this, and the previous, figure shows our model points also have the potential to explain the origin of the 130 GeV $\gamma$-ray line observed from the Galactic Centre in terms of the annihilation of a 130 GeV neutralino \cite{Tomar:2013zoa}.


\begin{figure}[t]
\centering
\includegraphics[width=0.75\linewidth]{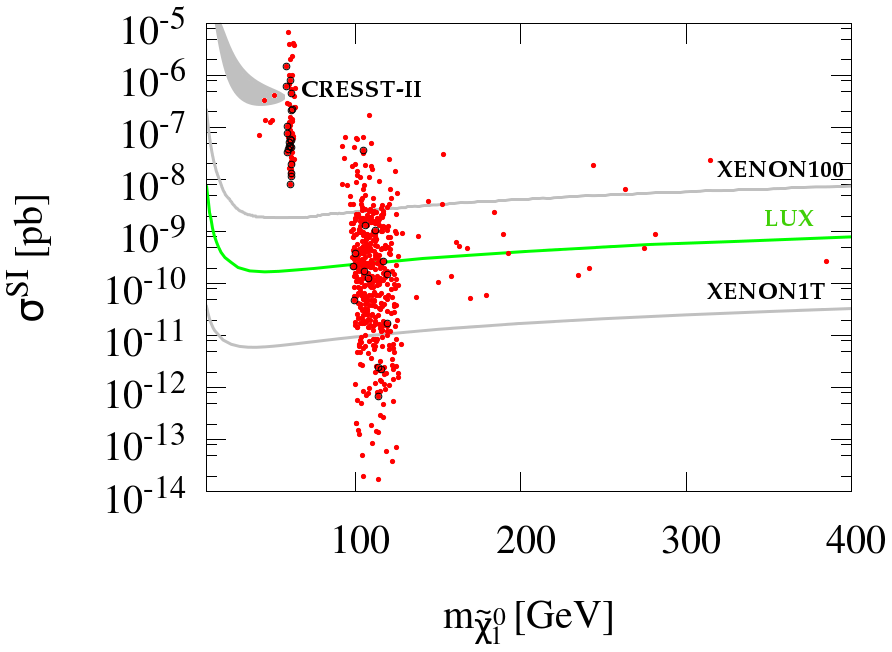}
\caption{Spin independent direct detection cross section vs. mass of the neutralino in our scan. 
The current bounds from XENON100 and the 2$\sigma$ signal region from CRESST--II are shown.
We also show projected bounds for XENON1T and LUX experiments. 
The points where neutralino relic abundance accounts for the full dark mattter content of the Universe measured by Planck within $3\sigma$, i.e. $0.1118<\Omega_{{\widetilde{\chi}}_1^0} h^2<0.128$, are highlighted in black circles.}
\label{sigma}
\end{figure}


In Fig.~\ref{sigma}, we show how spin independent dark matter--nucleon scattering experiments probe the scenarios with the constraints listed in Table \ref{tablecons} and $F(\gamma)=0$, see Eq.~(\ref{F-gamma-0}). 
The points that are circled in black fall within the range $0.1118<\Omega_{{\widetilde{\chi}}_1^0} h^2<0.128$ set by Planck. 
It is interesting to note that quite a few points lie in the regions where LSP is relatively light 
within the ranges where DAMA/LIBRA \cite{Bernabei:2008yi}, CRESST--II \cite{Angloher:2011uu}, CoGeNT \cite{Aalseth:2010vx} and CDMS \cite{Agnese:2013cvt} detected excess interactions in background. 
All the points with the smallest $m_{\widetilde{\chi}_1^0}$ have a large singlino fraction, and are ruled out by XENON100 experiment. 
We also show the projected bounds from the XENON1T and LUX experiments. 
These bounds are based on the assumptions already mentioned in the \ref{constraints app} section. 

Since in most of the cases that we have found the relic abundance is significantly lower than the set limits by the direct detection experiments,
we need to lower the cross section $\sigma^{SI}$ by a factor $(\Omega_{\widetilde{\chi}_1^0}/\Omega_{observed})$. 
Here we take the Planck central value $\Omega_{observed}=0.1199$.  As we see, most of the points that fall below XENON100 will be tested very soon by LUX and XENON1T experiments.






\begin{figure}[htbp]
    \centering
    \subfigure
    {
        \includegraphics[width=2.8in,height=1.99in]{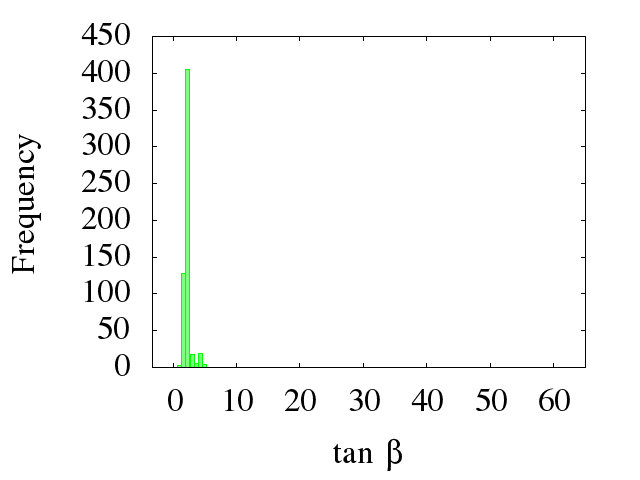}
        \label{fig:first_sub}
    }
    \subfigure
    {
        \includegraphics[width=2.84in,height=1.99in]{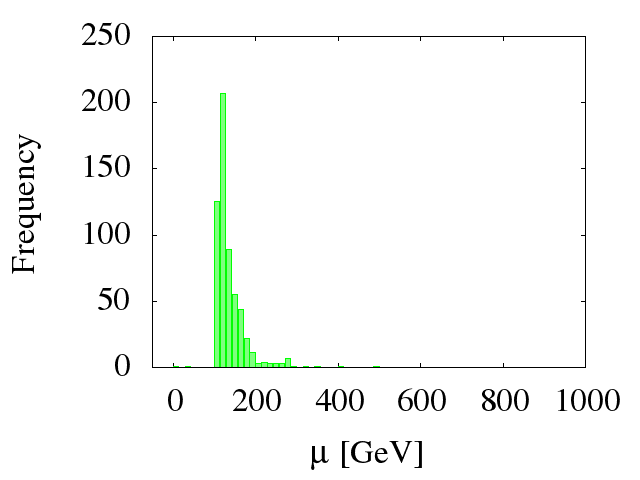}
        \label{fig:second_sub}
    }
\
    \subfigure
    {
        \includegraphics[width=2.84in,height=1.99in]{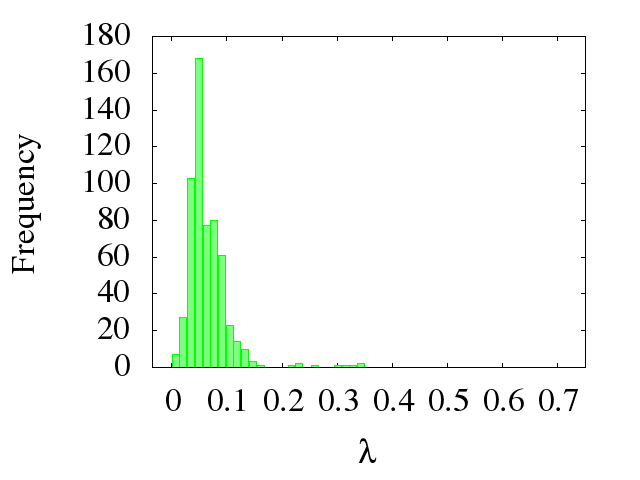}
        \label{fig:third_sub}
        
    }
    \subfigure
    {
        \includegraphics[width=2.84in,height=1.9in]{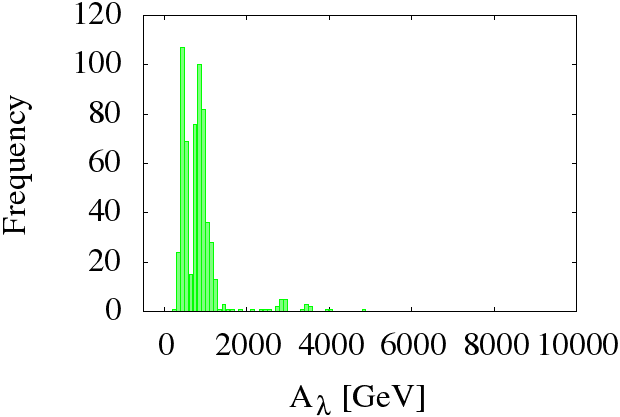}
        \label{fig:fourth_sub}
    }
    \caption{Distribution of the parameters which are relevant for baryogenesis in our scan.}
    \label{dist}
\end{figure}

In Fig.~\ref{dist}, we  show how the relevant parameters that enter  Eq.~(\ref{F-gamma-0}) are distributed in the scans. 
In the top left panel we can see that $\tan \beta$ tend to cluster around lower values. 
This is not because a higher $\tan \beta $ is inconsistent with the baryon washout condition, just that our approximations break down for the large $\tan \beta$ so we avoided scanning those. 
The breakdown is due to terms in $\Delta V$ that are $\tan \beta$ dependent and for large $\tan \beta$ can make $\Delta V$ too large so that our assumption of $\Delta V$ being small is violated. 
Similarly the upper bound on $A_\lambda$ and $\lambda$ is a relic of our approximations rather than any real difficulty in satisfying the baryon washout condition in that parameter range. 
The lower bound on $\lambda$, however, originates from baryogenesis since the low $\kappa$ and low $\lambda$ region is the MSSM limit and it is difficult to satisfy the baryon washout condition in the MSSM for a Higgs mass of $125$ GeV \cite{Carena:2012np}. 
We kept $A_t=-5000$ GeV fixed to be able to satisfy the Higgs mass bound more easily.
Values of $\mu$ are mainly within a 100--200 GeV range because, as mentioned above, large values are constrained by the requirement of $m_s^2$ being positive.  
This translates into upper bound, lower than $A_\lambda \cos\beta\sin\beta$, for a particular $\tan\beta$. 
Lower $\mu$ values are constrained because of the invisible Z decay and the chargino mass, which set bounds on the higgsino component of the neutralino which is directly related to low $\mu$.
Since $A_\lambda$ enters Eq. (\ref{equation}) through $\widetilde{a}$, in order to satisfy condition $F(\gamma)=0$ there needs to be some tuning between fourth and fifth terms.  
This fine tuning increases with increasing $A_\lambda$, and so condition \ref{equation} is much easier met at low values.
 

\section{Gauge invariant inflaton}
\label{inflationnn}
Within the MSSM there are nearly 300 gauge-invariant $F$- and $D$-flat directions, for a review see~\cite{Enqvist:2003gh}. Out of these flat directions, we will be interested in studying $\widetilde{u}\widetilde{d}\widetilde{d}$ as an inflaton~\cite{Allahverdi:2006iq,Allahverdi:2006we,Allahverdi:2006cx}, where $\widetilde u,~\widetilde d$ correspond to the right handed squarks. In fact within the MSSM, $\widetilde L\widetilde L \widetilde e$~\cite{Allahverdi:2006iq,Allahverdi:2006cx} and $H_u H_d$~\cite{Chatterjee:2011qr} could also be good inflaton candidates.  All the inflaton candidates provide {\it inflection point} in their respective potentials~\cite{Enqvist:2010vd}, where inflation can be driven for sufficiently large e-foldings of inflation to explain the current Universe and explain the seed perturbations for the temperature anisotropy in the CMB, which has been confirmed by the recent 
Planck data~\cite{Wang:2013hva,Choudhury:2013jya}. 
 
Within the NMSSM with the introduction of a singlet it becomes necessary to include the dynamics of a singlet field. Since the singlet here is not gauged there will be contributions to the $SH_uH_d$ potential which would potentially ruin the flatness and therefore the success of inflation driven solely by 
the gauge invariant inflaton~\cite{Mazumdar:2010sa}. While $\widetilde L\widetilde L\widetilde e$ could still be a good inflaton candidate in our current scenario, but in our analysis slepton masses are not constrained by the dark matter and the baryogenesis constraints, therefore we only concentrate on $\widetilde u\widetilde d\widetilde d$ as an inflaton for this study.  

Previous studies on MSSM inflation only considered the overlap between the parameter space for a successful inflation and the neutralino as the dark matter, which satisfies the relic abundance~\cite{Allahverdi:2007vy,Allahverdi:2010zp,Boehm:2012rh}. In this paper, we will consider one further step and we wish to constrain primordial inflation along with neutralino dark matter and condition for sufficient baryogenesis.

\subsection{Brief review of inflation}

The $\widetilde{u}\widetilde{d}\widetilde{d}$ flat direction is lifted by a higher order superpotential term of the following form \cite{Enqvist:2003gh},
\begin{equation} \label{eq-i-supot}
W \supset {y \over 6}{\Phi^6 \over M^3_{p}}\, ,
\end{equation}
where $y \sim {\cal O}(1)$, and the scalar component of the superfield $\Phi$ is given by:
\begin{equation} \label{eq-i-infl}
\phi = {\widetilde{u} + \widetilde{d} + \widetilde{d} \over \sqrt{3}} \,.
\end{equation}
After minimizing the potential along the angular direction $\theta$ ($\Phi$ = $\phi e^{i \theta}$), we can consider the real part of $\phi$
, for which the scalar potential is then given by~\cite{Allahverdi:2006iq,Allahverdi:2006we}
\begin{equation} \label{eq-i-scpot}
V(\phi) = {1\over2} m^2_\phi\, \phi^2 - A_y {\phi^6 \over 6\,M^{6}_{p}} + 
{{\phi}^{10} \over M^{6}_{p}}\,,
\end{equation}
where $m_\phi$ and $A$ are the soft breaking mass and the $A_y$-term respectively ($A_y$ is a positive quantity since its phase is absorbed by a redefinition of $\theta$ during the process). The mass for $\widetilde{u}\widetilde{d}\widetilde{d}$ is given by:
\begin{eqnarray}\label{eq-i-masses}
m^2_{\phi}=\frac{m^2_{\widetilde u}+m^2_{\widetilde d}+m^2_{\widetilde d}}{3}\,.
\end{eqnarray}
Note that the masses are now VEV dependent, i.e. $m^2(\phi)$. The inflationary perturbations will be able to constrain 
the inflaton mass only at the scale of inflation, i.e. $\phi_0$, while LHC will be able to constrain the masses at the LHC scale. However both the physical 
quantities are related to each other via renormalization group equations (RGEs), as we shall discuss below.

For~\cite{Allahverdi:2006we}
\begin{equation} \label{eq-i-dev}
{A_y^2 \over 40 m^2_{\phi}} \equiv 1 + 4 \alpha^2\, ,
\end{equation}
where $\alpha^2 \ll 1$, there exists a point of inflection ($\phi_0$) in $V(\phi)$, where
\begin{eqnarray}
&&\phi_0 = \left({m_\phi M^{3}_{p}\over \lambda \sqrt{10}}\right)^{1/4} + {\cal O}(\alpha^2) \, , \label{eq-i-infvev} \\
&&\, \nonumber \\
&&V^{\prime \prime}(\phi_0) = 0 \, , \label{eq-i-2nd}
\end{eqnarray}
at which
\begin{eqnarray}
\label{eq-i-pot}
&&V(\phi_0) = \frac{4}{15}m_{\phi}^2\phi_0^2 + {\cal O}(\alpha^2) \, , \\
\label{eq-i-1st}
&&V'(\phi_0) = 4 \alpha^2 m^2_{\phi} \phi_0 \, + {\cal O}(\alpha^4) \, , \\
\label{eq-i-3rd}
&&V^{\prime \prime \prime}(\phi_0) = 32\frac{m_{\phi}^2}{\phi_0} + {\cal O}(\alpha^2) \, .
\end{eqnarray}
The Hubble expansion rate during inflation is given by~\cite{Allahverdi:2006iq,Allahverdi:2006we}
\begin{equation}\label{eq-i-hubble}
H_{inf} \simeq \frac{1}{\sqrt{45}}\frac{m_{\phi}\phi_0}{M_{p}}\,.
\end{equation}

The amplitude of the initial perturbations and the spectral tilt are given by:
\begin{equation} \label{eq-i-ampl}
\delta_H = {8 \over \sqrt{5} \pi} {m_{\phi} M_{p} \over \phi^2_0}{1 \over \Delta^2}
~ {\rm sin}^2 [{\cal N}_{\rm COBE}\sqrt{\Delta^2}]\,, \end{equation}
and
\begin{equation} \label{eq-i-tilt}
n_s = 1 - 4 \sqrt{\Delta^2} ~ {\rm cot} [{\cal N}_{\rm COBE}\sqrt{\Delta^2}], \end{equation}
respectively, where
\begin{equation} \label{eq-i-Delta}
\Delta^2 \equiv 900 \alpha^2 {\cal
N}^{-2}_{\rm COBE} \Big({M_{p} \over \phi_0}\Big)^4\,. \end{equation}
In the above, ${\cal N}_{\rm COBE}$ is the number of e-foldings between the time when the observationally relevant perturbations are generated till the end of inflation and follows: ${\cal N}_{\rm COBE} \simeq 66.9 + (1/4) {\rm ln}({V(\phi_0)/ M^4_{p}}) \sim 50$. Since inflaton is made up of squarks, reheating and thermalisation happens instantly as showed in Ref.~\cite{Allahverdi:2011aj}. The reheat temperature at which all the {\it degrees of freedom} are in thermal equilibrium (kinetic and chemical equilibrium) is given by
$T_{rh}=\left({120}/{15\pi^2g_\ast}\right)^{1/4}\sqrt{m_\phi \phi_0} $,
where $g_\ast=228.75$ is the relativistic degrees of freedom present within MSSM.
Since all the physical parameters are fixed in this model once we determine the soft SUSY breaking mass, $m_\phi$, the estimation of the
reheat temperature can be made rather accurately. 

\begin{figure}[t]
\centering
\includegraphics[width=0.75\linewidth]{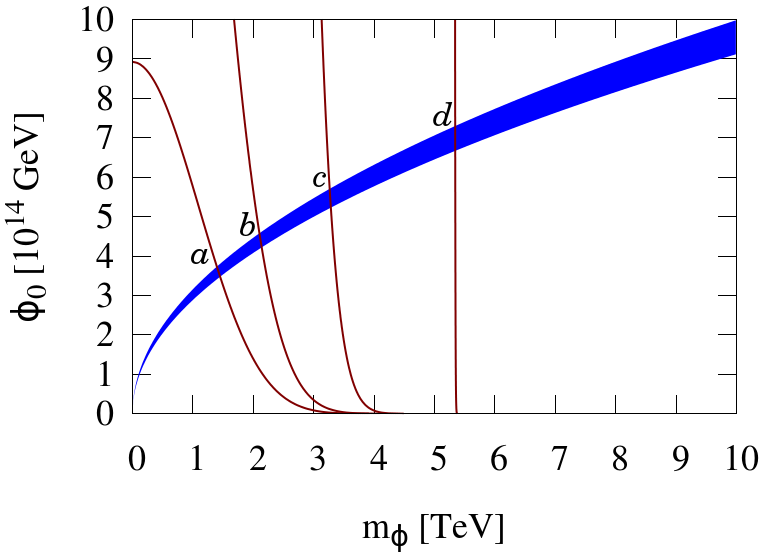}
\caption{Blue region depicts the parameter space for inflation
where it yields the right amplitude of density perturbations in the CMB, i.e. $P_\zeta=2.196\times 10^{-9}$ and the $\pm 1\sigma$ variance of the spectral tilt, $n_s=0.9606\pm0.0073$. The brown lines show the running of the inflaton mass, where they intersect with the blue region depict the
correct relic abundance,  $0.1118<\Omega_{{\widetilde{\chi}}_1^0} h^2<0.128$ \cite{Ade:2013zuv}, and strongly first order phase transition.
From these intersections, $a, b, c, d$ we can determine the masses of the inflaton at the inflationary scale $\phi_0$. The running of the inflaton mass is mainly determined by the bino and gluino masses, see Table \ref{tab:summary}. }
\label{inflation}
\end{figure}


\begin{table}
\begin{center}
\begin{tabular*}{1.05\textwidth}{c|c|c|c|c|c}
\hline 
Parameter & a & b & c & d & Constrained by \\ 
\hline 
\hline 
       $\lambda$  & $0.06405$ & $0.06520$ & $0.10418$ & $0.02906$  &$1^{st}$ order phase transition\\ 
       $A_\lambda ({\rm GeV})$  & $1127$ & $524$ & $772$ & $796$  &$1^{st}$ order phase transition \\ 
       $\tan\beta$  & $2.659$ & $2.042$ & $2.276$ & $2.123$  &$1^{st}$ order phase transition\\ 
       $\mu_\text{eff} ({\rm GeV})$  & $165.214$ & $142.985$ & $176.515$ & $137.963$  &$1^{st}$ order phase transition\\ 
       $\Omega_{\widetilde{\chi}_1^0} h^2$  & $0.119$ & $0.112$ & $0.124$ & $0.112$  &Dark Matter abundance\\ 
       $m_{\widetilde{\chi}^0_1({\rm GeV})}$  & $61.17$ & $119.2$ & $59.8$ & $126.3$  &Dark Matter abundance\\ 
       $M_1 ({\rm GeV})$  & $2151$ & $2006$ & $1375$ & $1084$ & Inflaton RGE\\ 
       $M_3 ({\rm GeV})$  & $5269$ & $4986$ & $4281$ & $861$ &Inflaton RGE \\ 
       $\phi_0[\times10^{14}] ({\rm GeV})$  & $3.5-3.8$ & $4.2-4.6$ & $5.3-5.7$ & $6.6-7.3$  &CMB temperature anisotropy\\ 
       $m_\phi ({\rm GeV})$  & $1425$ & $2120$ & $3279$ & $5349$  &CMB temperature anisotropy  \\ \hline\hline
\end{tabular*}
\caption{\label{tab:summary-1}We show the benchmark points that are depicted in Fig.~\ref{inflation}. The gaugino masses which enter in the RG equations are mainly sensitive due to different $M_1$ and $M_3$. The parameters $\lambda, A_\lambda, \tan\beta, \mu_\text{eff}$ are constrained from baryogenesis point of view, and this in turn uniquely determine the mass of the lightest stop which sets the mass for $\widetilde{u}_3\widetilde{d}_i\widetilde{d}_j$ inflaton candidate ($3\neq i\neq j$). Once again we reiterate that without our approximation scheme, the constraints on baryogenesis would be significantly less strict. The mass of the inflaton is given at the inflationary scale $\phi_0$. }
\end{center}
\end{table}


\subsection{Parameter space for inflation, dark matter and baryogenesis}
Since the requirement for a successful baryogenesis implicitly constraints the right handed squark, i.e. $\widetilde u_3$, we can assign 
the flat direction combination to be: $\widetilde {u}_i\widetilde {d}_j\widetilde {d}_k$, where  $i=3$ and $i\neq j\neq k$. 

We can then use the RGEs for the $\widetilde u\widetilde d\widetilde d$ flat direction, in order to relate the low energy physics that we can 
probe at the LHC with the high energy inflation, which is constrained by the Planck data. The RGEs for the inflaton mass and the $A_y$-term are given by~\cite{Allahverdi:2006we}:
\begin{equation}
\begin{aligned}
\label{rgudd}
&\hat{\mu} \frac{dm^2_\phi}{d\hat{\mu}}=-\frac{1}{6\pi^2}\bigg(4M_3^2g_3^2+\frac{2}{5}M_1^2g_1^2\bigg),
\\&\hat{\mu} \frac{dA_y}{d\hat{\mu}}=-\frac{1}{4\pi^2}\bigg(\frac{16}{3}M_3g_3^2+\frac{8}{5}M_1g_1^2\bigg),
\end{aligned}
\end{equation}
where  $\hat{\mu} = \hat\mu_0=\phi_0$ is the VEV at which inflation occurs,
$M_1$ and $M_3$ are $U(1)$ and $SU(3)$ gaugino masses,  and $g_1$ and $g_3$ are the associated couplings. 

To solve these equations, we need to take into account of the running of 
the gaugino masses and coupling constants which are given by, see~\cite{Allahverdi:2006we}:
\begin{equation}
\beta (g_i)=\alpha_i g_i^3 \hspace{1.5cm}
\beta\bigg{(}\frac{M_i}{g_i^2}\bigg{)}=0,
\end{equation}
with $\alpha_1={11}/{16\pi^2}$ and $\alpha_3=-{3}/{16\pi^2}$.
Since from Eq.~(\ref{eq-i-masses}), we know the mass of the inflaton at the electroweak scale, by using the RGEs 
we are able to evolve it to the high scale $\phi_0$, where inflation can happen. 
This can be seen in Fig.~\ref{inflation}. The blue region shows the parameter 
of $\widetilde u_3\widetilde d_j\widetilde d_k$ as an inflaton for $j\neq k\neq 3$. 
It includes central value of density  perturbations together with $\pm 1\sigma$ variation in spectral tilt $n_s$. 
The brown lines show the mass of the inflaton at a particular scale and its running from high scale to low scale 
is determined by the RGEs, and it is mostly sensitive to bino and gluino masses.

In  Fig.~\ref{inflation}, we show  the four benchmark points, $a,~b,~c,~d$, which satisfy the condition for a successful baryogenesis, Eq.~(\ref{F-gamma-0}), and also accommodate neutralino as a dark matter which satisfies the relic abundance constraint $0.1118<\Omega_{{\widetilde{\chi}}_1^0} h^2<0.128$.
In table \ref{tab:summary-1}, we summarise the relevant parameters for NMSSM required to explain the Universe beyond the 
Standard Model.

\section{Conclusions}

In this work we examined inflation, baryogenesis and dark matter in the context of Next-to-Minimal Supersymmetric Standard Model.  We have found that these three important cosmological requirements can be simultaneously accommodated by the theory.  In particular, we have shown that a strongly first order phase transition can be easily achieved even with recent LHC constrains applied.  Then we demonstrated that an abundance of lightest neutralinos can be generated thermally which satisfies the present dark matter density limits.  Part of these model points also pass the most stringent dark matter direct detection constraints.  Finally, we have shown that the presented scenario is fully consistent with inflation, where the inflaton is a $D$-flat direction and it is made up of right handed squarks.  The visible sector inflation would explain not only the temperature anisotropy of the CMB, but also all the relevant matter required for baryogeneis and dark matter.


\begin{acknowledgments}

We thank Michael Morgan for invaluable help, continuous discussions, feedback and support throughout this project. EP also would like to thank Sofiane Boucenna, Jonathan Da Silva and Lingfei Wang for valuable discussions. This work was supported in part by the {\it ARC Centre of Excellence for Particle Physics at the Tera-scale}. EP is supported by STFC ST/J501074. AM is supported by the Lancaster-Manchester-Sheffield Consortium for Fundamental Physics under STFC grant ST/J000418/1.  CB thanks the Kavli Institute of Theoretical Physics, China for providing partial financial support and an excellent research environment during part of this project.

\end{acknowledgments}


\section{Appendix: Solving the toy model}
The first derivative with respect to $\varphi _c^2$ is 
\begin{eqnarray} \left. \varphi ^2 \frac{\partial V}{\partial \varphi ^2}\right| _{\varphi _c, T_c} &=& M^2(T)\varphi _c^2 - \frac{3T_cE}{2}\varphi ^3 + \tilde{\lambda} \varphi _c^4 \nonumber \\ &-& \frac{ \tilde{a}^2 \varphi^4 (2m_s^2 + \lambda ^2 \varphi _c^2)}{(m_s^2+\lambda ^2 \varphi ^2)^2} \label{firstdir} \ . \end{eqnarray}

Using the condition that the potential at the critical VEV is equal to the potential at $\varphi=0$ we have a second equation
\begin{equation} M^2(T_c) \varphi _c^2 - T_cE\varphi _c^3 + \frac{\tilde{\lambda }}{2}\varphi _c^4 -\frac{\tilde{a}^2\varphi_c^4}{m_s^2+\lambda \varphi _c^2} =0  \label{gcond2} \ . \end{equation}
We can then use Eq. (\ref{firstdir}) and set it to $3/2$ times Eq. (\ref{gcond2}) to get 
\begin{equation} cT_c ^2 + M^2 = \frac{\varphi _c^2}{2}\left( \tilde{\lambda } - \frac{2\tilde{a}^2 m_s^2}{(m_s^2+\lambda^2\varphi_c^2)^2} \right) + \frac{\lambda ^2 \tilde{a}^2 \varphi _c^4}{(m_s^2+ \lambda ^2 \varphi _c^2)^2} \label{treeeq2} \end{equation}
Finally we can set Eq. (\ref{firstdir}) equal to Eq. (\ref{gcond2}) to get a second equation 
\begin{equation} E T_c \varphi _c^3 = \tilde{\lambda} \varphi _c^4 - \frac{2 \tilde{a}^2 m_s^2 \varphi _c^4}{(m_s^2 + \lambda ^2 \varphi _c^2)^2} \ . \label{treeeq1} \end{equation}
We can then divide both sides of the above equation to obtain
\begin{equation} E \gamma = \tilde{ \lambda} - \frac{2 \tilde{a}^2 m_s^2}{(m_s^2+\lambda ^2\varphi _c^2)^2} \ . \end{equation}
To solve this we use the ansatz $\varphi _c^2 = \frac{1}{\lambda ^2}(-m_s^2 + \delta)$ and it is straight forward to show that
\begin{equation} \varphi _c^2 = \frac{1}{\lambda^2} \left( -m_s^2 + \sqrt{2\frac{\tilde{a}^2m_s^2}{\tilde{\lambda} - \gamma E}} \right) \label{phisol} \ .  \end{equation}
Consider the first term on the right hand side of Eq. (\ref{treeeq2}). It is proportional to $\gamma$. Using this and dividing both sides by $\varphi _c^2$ we have 
\begin{equation} c \gamma ^2= \frac{G}{\varphi _c^2}+\frac{\gamma E}{2}+\frac{\lambda ^2 \tilde{a}^2 \varphi _c^2}{(m_s^2 + \lambda ^2 \varphi _c^2 )^2} \ . \end{equation}
We then use Eq. (\ref{phisol}) to get the following equation
\begin{eqnarray}0 &=&-\frac{\tilde{\lambda}}{2}+\gamma E-c\gamma^2+ \frac{\sqrt{\tilde{a}^2(\tilde{\lambda}-\gamma E)}}{ \sqrt{2} m_s} + \frac{\lambda^2 G}{-m_s^2 + \sqrt{\frac{2 \tilde{a}^2 m_s^2}{\tilde{\lambda}-\gamma E}}}  \nonumber \\ & \equiv & F(\gamma) \ \label{equation} \end{eqnarray} 
as required.

\end{document}